\documentclass[letterpaper, 10 pt, conference]{ieeeconf}  

\IEEEoverridecommandlockouts                              

\usepackage{graphics} 
\usepackage{epsfig} 
\usepackage{mathptmx} 
\usepackage{times} 
\usepackage{amsmath} 
\usepackage{amssymb}  

\makeatletter
\def\endtable{\end@float}
\makeatother
\usepackage{subfig}
\usepackage{booktabs}

\usepackage{eso-pic} 

\title{\LARGE \bf
Characterization of Speech Imagery in Scalp EEG and Comparison with Motor Imagery
}

\author{
Bob Van Dyck$^{1}$, Liuyin Yang$^{1}$, Qiang Sun$^{1}$,
Ang Li$^{2}$, Marc M. Van Hulle$^{1}$%
\thanks{
  $^{1}$Laboratory for Neuro- \& Psychophysiology, Department of Neurosciences, KU Leuven, B-3000 Leuven, Belgium. Corresponding author: Bob Van Dyck (e-mail: bob.vandyck@kuleuven.be).
}
\thanks{
  $^{2}$Shanghai Advanced Research Institute, Chinese Academy of Sciences, Shanghai, China.
  }
}

\begin{document}

\AddToShipoutPictureFG*{%
  \AtPageLowerLeft{%
    \raisebox{0.28in}{%
      \makebox[\paperwidth]{%
        \parbox{0.90\paperwidth}{%
          \centering\footnotesize
          This work has been submitted to the IEEE for possible publication. Copyright may be transferred without notice, after which this version may no longer be accessible.
        }%
      }%
    }%
  }%
}

\maketitle
\thispagestyle{empty}
\pagestyle{empty}


\begin{abstract}

Speech imagery is an attractive brain-computer interface paradigm for communication because it is endogenous and intrinsically linguistic. 
Yet despite growing interest, its dominant scalp-EEG spatiotemporal characteristics remain poorly characterized. 
We investigated whether speech imagery, understood here as the motor imagery of articulatory movements, exhibits the motor-related mu/alpha and beta desynchronization expected from motor imagery. 
In $34$ participants, we compared speech imagery, finger motor imagery, and explicitly cued no-task trials recorded under the same trial structure, analyzing band-power dynamics across channels and time. 
Finger motor imagery showed the expected contralateral mu/alpha and beta desynchronization over sensorimotor areas, whereas speech imagery showed a weaker, more distributed increase in alpha power relative to no-task. 
A classifier discriminating imagery from no-task reached mean balanced accuracies of $0.563 \pm 0.071$ for speech imagery and $0.717 \pm 0.125$ for motor imagery, with band-ablation analyses showing larger, more robust alpha and beta effects for motor imagery. 
These results show that the dominant group-level scalp response to speech imagery did not resemble the canonical alpha/beta desynchronization associated with motor imagery. 

\end{abstract}


\section{INTRODUCTION}

Speech imagery is an attractive brain-computer interface paradigm for communication because it is endogenous and intrinsically linguistic. This distinguishes it from reactive paradigms such as P300 and SSVEP spellers, which depend on external stimulation, and from motor imagery BCIs, which are endogenous but typically rely on non-linguistic surrogate control signals \cite{Nguyen2018,LopezBernal2022}. Although several studies have decoded speech imagery from scalp EEG, performance varies widely across datasets, preprocessing choices, and decoding strategies \cite{LopezBernal2022,Panachakel2021}. Despite these encouraging results, the dominant group-level scalp-EEG signature of speech imagery remains poorly characterized.

Following Schultz et al. \cite{Schultz2017}, we use \emph{speech imagery} to denote the motor imagery of speaking without overt articulator movement, distinct from \emph{inner speech} as self-directed verbal thought. Speech imagery is often conceptualized as an internal simulation process linking motor and perceptual systems \cite{TianPoeppel2012,TianPoeppel2013,TianZaratePoeppel2016}. Intracranial recordings suggest a shared spectral signature across manual and articulatory movements, as both executed hand and tongue movements are accompanied by alpha- and beta-range attenuation \cite{crone1998alpha,miller2007spectral}. Low-beta decreases have also been observed during imagined speech, although these effects are weaker and occur at fewer sites than during overt speech \cite{ProixEtAl2022}.

Using finger motor imagery as a reference condition, we asked whether speech imagery produces motor-related mu/alpha or beta desynchronization at the scalp, potentially with a topography distinct from the typical contralateral central sensorimotor pattern classically associated with hand and finger motor imagery \cite{PfurtschellerLopesdaSilva1999,Ramoser2000}, or whether its dominant response instead reflects broader task-related activity. To address this question, we used a matched within-participant multi-task EEG dataset comprising speech imagery, finger motor imagery, and explicitly cued no-task trials under the same timing structure. We characterized the dominant group-level speech-imagery pattern and assessed whether the spatiotemporal and decoding results supported a motor-related interpretation.

\section{METHODS}

\subsection{Dataset, task design, and preprocessing}

\begin{figure}[t]
    \centering
    \includegraphics[width=\columnwidth]{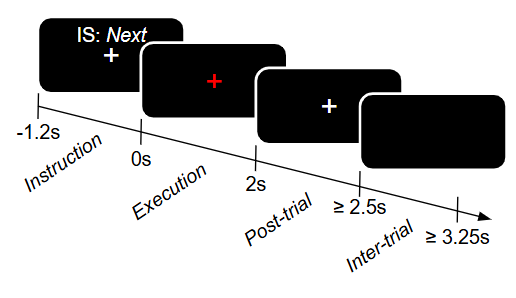}
    \caption{Trial structure consisting of a $1.2$~s instruction period, a $2$~s execution period, and jittered post- and inter-trial intervals.}
    \label{fig:trial_structure}
\end{figure}

We analyzed data from $34$ healthy adults recruited as right-handed ($18$ female participants, aged $20$--$33$~years) from a multitask EEG dataset \cite{Yang2026Dataset}. Handedness was assessed using the Edinburgh Handedness Inventory \cite{Oldfield1971}, with $30$ of $34$ participants showing laterality scores above $70$. EEG was recorded with a Neuroscan SynAmps RT system using $64$ scalp EEG channels and four electrooculography channels at a sampling rate of $500$~Hz. 
Participants completed multiple BCI tasks within a single session, including imagined speech and imagined finger movements. The imagined-speech task comprised three navigation commands (\emph{next}, \emph{back}, \emph{select}) and a \emph{no-task} condition. Participants were instructed to imagine speaking the cued command by focusing on the articulatory movements involved, without executing them or producing sound. The imagined-finger movement task used three finger flexions (\emph{thumb}, \emph{index}, \emph{point}), where point denotes flexion of all fingers except the thumb and index, also with a \emph{no-task} condition. Participants were instructed to imagine performing the cued flexion without executing the movement. In both tasks, participants were instructed to maintain fixation and refrain from speech or movement imagery during the \emph{no-task} condition.
Trials consisted of a $1.2$~s instruction period, followed by a $2$~s execution period marked by a change in fixation color at $0$~s, and jittered post-trial ($0.5$ to $0.7$~s) and inter-trial ($0.75$ to $1.5$~s) intervals (Fig.~\ref{fig:trial_structure}). 
Each participant completed $60$ trials per imagery target and $60$ no-task trials in both the imagined-speech and imagined-finger-movement tasks.
Speech- and motor-imagery trials were acquired in separate blocks. Within each block, imagery targets and no-task trials were presented in randomized order, and the sequence of speech- and motor-imagery blocks was randomized for each participant.
The study was approved by the Ethical Committee of UZ Leuven (S62547), and all participants provided written informed consent. 

Preprocessing included notch filtering, $0.5$ to $100$~Hz band-pass filtering, bad-channel interpolation, ICA-based artifact removal, and downsampling to $200$~Hz. For the present analysis, epochs were extracted from $-1.5$ to $3.0$~s relative to the go cue. We computed Welch power spectral density in the delta ($1$--$4$~Hz), theta ($4$--$8$~Hz), alpha ($8$--$12$~Hz), low beta ($12$--$18$~Hz), high beta ($18$--$30$~Hz), low gamma ($30$--$45$~Hz), and high gamma ($45$--$90$~Hz) bands using non-overlapping $0.5$~s windows. Trial-level power estimates were aggregated using the geometric mean and subsequently log-transformed.
For a descriptive baseline-normalized comparison, mean log-power from a pre-instruction baseline interval ($-2.0$ to $-1.5$~s) was subtracted from the analyzed time courses separately for each participant, condition, channel, and frequency band.

\subsection{Spatiotemporal cluster-permutation analysis} 

For speech imagery, the three speech targets were averaged to form a single speech condition, and the corresponding contrast was computed by subtracting no-task. For motor imagery, the three finger-movement targets were averaged analogously, and the contrast was computed by subtracting no-task.
Statistical significance was assessed separately within each frequency band using a spatiotemporal cluster-based permutation test on participant-wise task-versus-no-task differences, with $10{,}000$ permutations and Holm-Bonferroni correction across the displayed bands \cite{Maris2007,Holm1979,Durka2004}. As usual for cluster-based inference, significance is interpreted at the level of extended effects rather than exact onsets or precise locations \cite{Pernet2015,Sassenhagen2019,Rousselet2025}. 
For a descriptive comparison, we additionally examined baseline-normalized speech-imagery activity without no-task subtraction, to assess whether its dominant pattern was consistent with the task-versus-no-task contrast.

To assess whether pooling across targets obscured target-specific patterns, we compared the three target-specific task-versus-no-task maps within each task using a repeated-measures spatiotemporal cluster-permutation test, separately for each frequency band, with $10{,}000$ within-participant permutations of target labels and Holm correction across bands. 
In addition, for each significant primary task-versus-no-task cluster, target-specific contrast values were averaged within the cluster and compared using a Friedman test, with Holm correction across clusters. 

\subsection{Filterbank Common Spatial Pattern for imagery detection} 

To relate the spatiotemporal results to imagery detection, we trained participant-specific filterbank Common Spatial Pattern (FB-CSP) detectors with shrinkage LDA on preprocessed epochs from $0$ to $2$~s after cue onset \cite{Ramoser2000,Ang2012}. The default model used six functional bands: delta, theta, alpha, beta, low gamma, and high gamma. Each epoch was decomposed into these bands, CSP filters were learned per band using shrinkage covariance regularization, and band-wise CSP power features were concatenated for classification. For each task, we trained three pairwise binary detectors, each contrasting one imagery target against the matched no-task condition. Performance was evaluated with $5$-fold cross-validation and summarized per participant by averaging balanced accuracy across the five folds and the three target-versus-no-task pairs within each task. Band contributions were probed with leave-one-band-out variants.
For descriptive heterogeneity analyses, participants were ranked by FB-CSP speech-imagery detection performance. We examined progressively smaller top-$k$ subsets and used an uncorrected participant-level binomial threshold to define the best-performing subset.
To assess target-specific decoding, we additionally applied the default six-band FB-CSP model to the three pairwise comparisons among the imagery targets within each task. 

\section{RESULTS}

\subsection{Speech imagery shows weak, distributed alpha synchronization}

\begin{figure*}[!t]
    \centering
    \subfloat[Speech imagery.\label{fig:speech_imagery_contrast}]{%
        \includegraphics[width=0.72\textwidth]{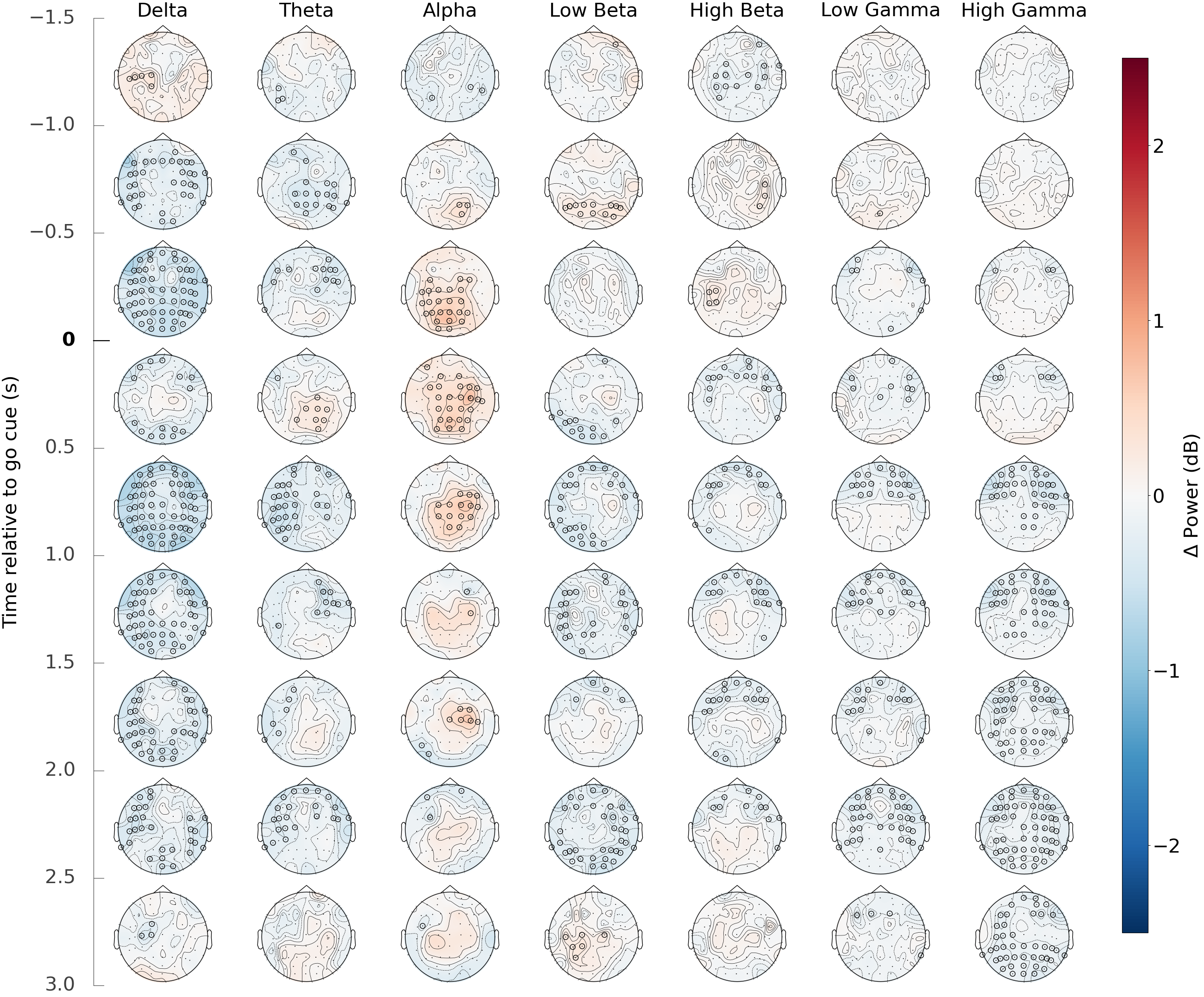}%
    }\\[1em]
    \subfloat[Motor imagery.\label{fig:motor_imagery_contrast}]{%
        \includegraphics[width=0.72\textwidth]{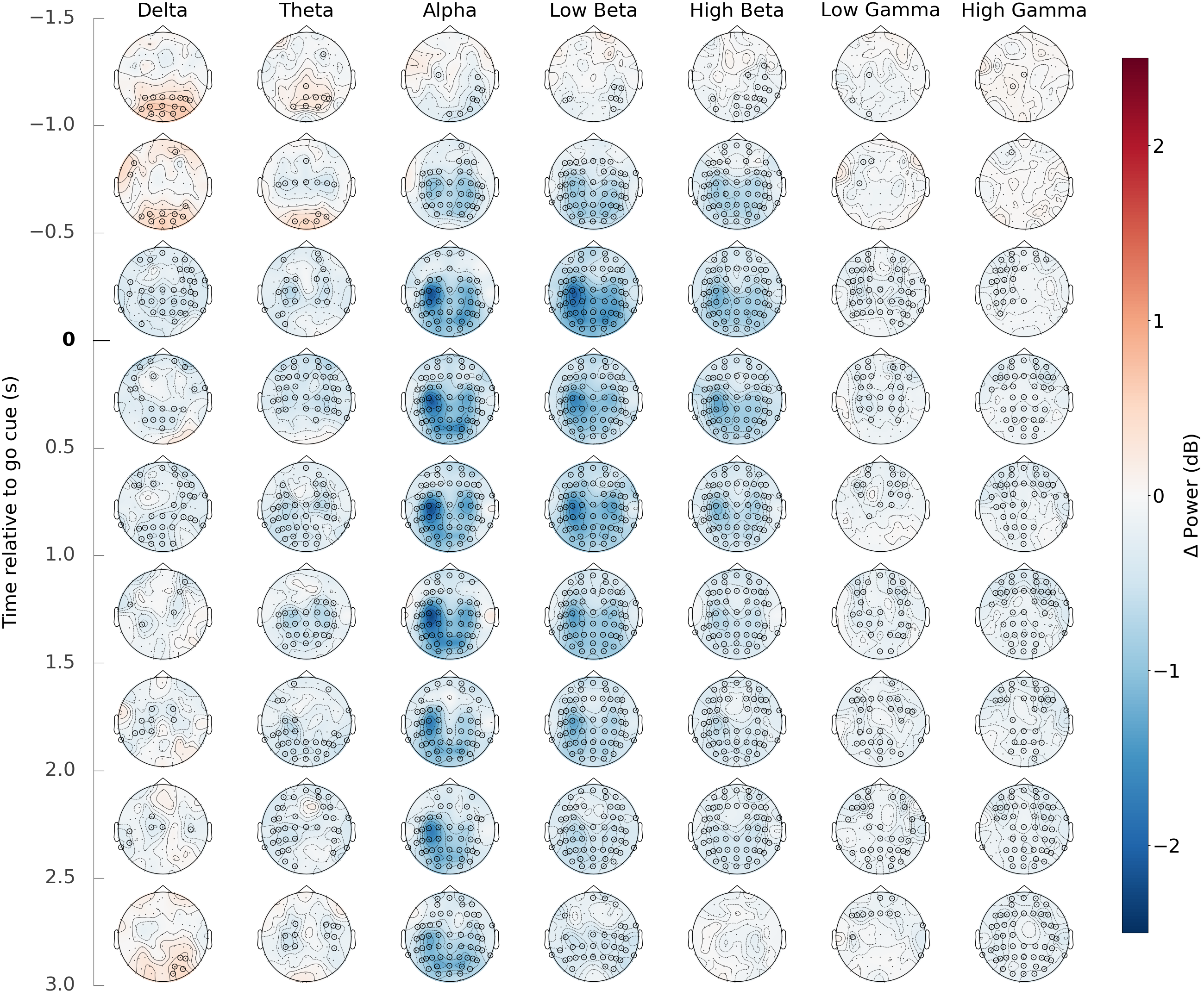}%
    }
    \caption{Group-level task-versus-no-task contrasts for speech and finger motor imagery. Open circles mark sensors in significant clusters after per-band cluster-permutation testing with Holm-Bonferroni correction across the seven displayed bands.}
    \label{fig:task_vs_notask}
\end{figure*}

Figure~\ref{fig:task_vs_notask} shows clearly different group-level task-versus-no-task contrasts for speech and motor imagery. Motor imagery produced the expected focal decrease in alpha and low-beta power over the left sensorimotor area, consistent with canonical contralateral event-related desynchronization during right-hand imagery \cite{PfurtschellerLopesdaSilva1999,Ramoser2000}. This desynchronization was sustained across the execution period and remained visible into the post-trial window. 
Speech imagery, by contrast, was characterized by a weaker and more spatially distributed alpha synchronization, strongest near the start of execution and not sustained across the full trial. 
Pre-instruction baseline normalization preserved a qualitatively similar distributed alpha pattern in the speech-versus-no-task contrast, but no alpha cluster remained significant after correction. 
A descriptive analysis of baseline-normalized speech-imagery trials without no-task subtraction retained the alpha synchronization and a significant alpha cluster, although with lower amplitude and an altered spatial distribution. Additional spectral and spatial changes were also evident across the lower-frequency bands.

\subsection{Speech imagery is harder to detect than motor imagery}

\begin{figure}[!t]
    \centering
    \includegraphics[width=0.8\columnwidth]{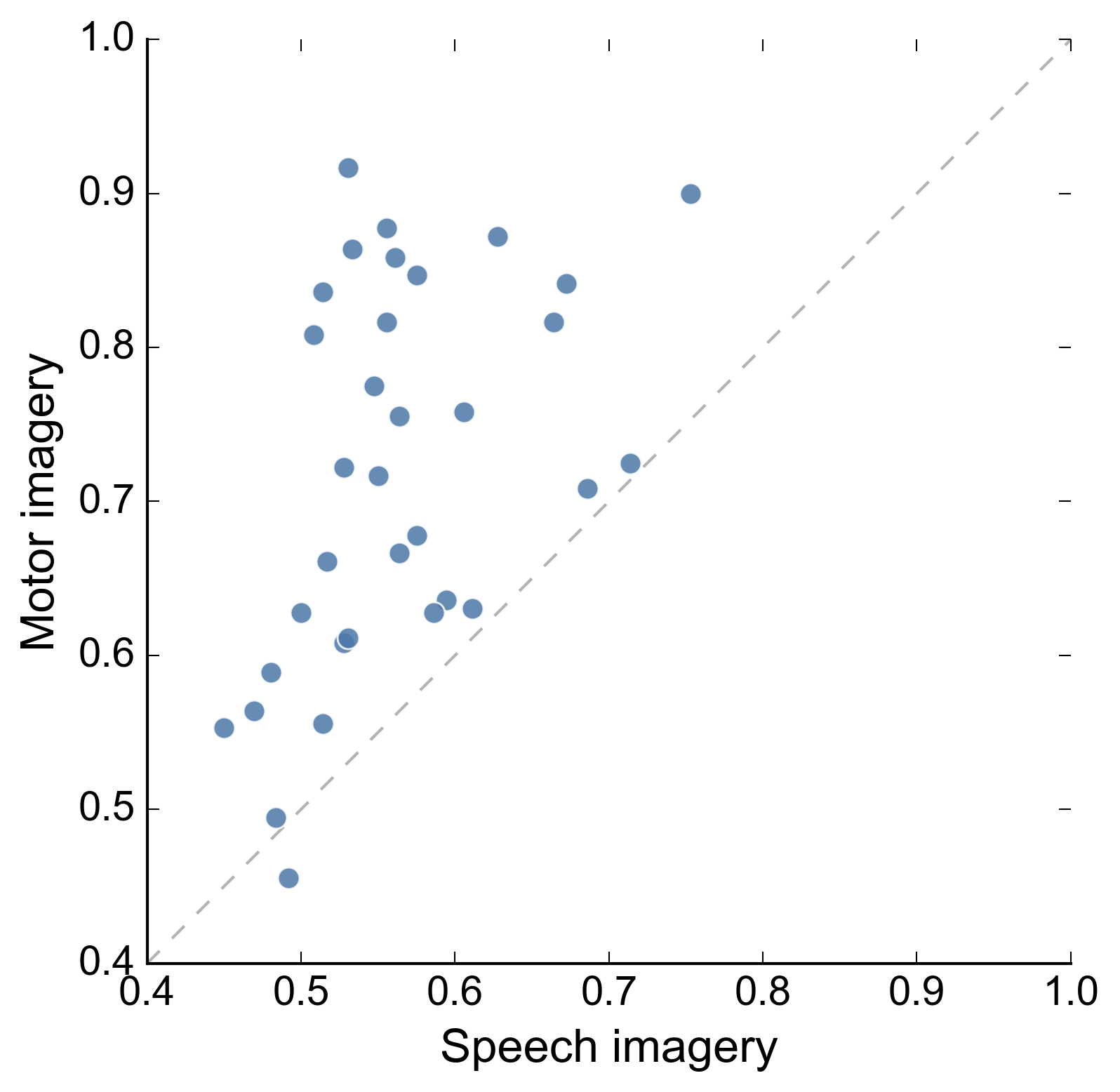}
    \caption{Participant-wise balanced accuracy for FB-CSP detection of speech and motor imagery.}
    \label{fig:speech_vs_motor_imagery_performance}
\end{figure}

Imagery detection with an FB-CSP classifier was comparatively straightforward for motor imagery, but weaker and less reliable for speech imagery. Across $34$ participants, mean balanced accuracy was $0.563 \pm 0.071$ for speech-versus-no-task detection and $0.717 \pm 0.125$ for motor-versus-no-task detection. As shown in Figure~\ref{fig:speech_vs_motor_imagery_performance}, participant-wise detection performance was moderately associated across tasks (Spearman's $\rho = 0.520$, $p = 0.00164$, $n = 34$). 

The leave-one-band-out results were broadly consistent with the spatiotemporal analysis (Table~\ref{tab:ablation}). Removing either alpha or beta significantly reduced detection performance for both tasks, with larger and more robust ablation effects for motor imagery. This stronger dependence matches the pronounced alpha and low-beta desynchronization observed over sensorimotor areas during motor imagery. For speech imagery, the larger alpha effect is consistent with its alpha-dominant scalp pattern, while the smaller but significant beta effect suggests additional discriminative information that was not expressed as a significant group-level spatiotemporal cluster.

\begin{table}[t]
\caption{Leave-one-band-out ablation study for imagery detection using FB-CSP. Default model performance is reported as mean balanced accuracy $\pm \mathrm{SD}$. Ablation values show the change after removing the indicated frequency band. Asterisks denote significant decreases based on paired Wilcoxon signed-rank tests across participants, with Holm correction across model configurations within each task ($^{*}p < 0.05$, $^{**}p < 0.01$, $^{***}p < 0.001$).} 
\label{tab:ablation}
\centering
\scriptsize
\setlength{\tabcolsep}{5pt}
\begin{tabular}{lll}
\toprule
& Speech imagery & Motor imagery \\
\midrule
\textit{Default model} & $0.563 \pm 0.071$ & $0.717 \pm 0.125$ \\
\midrule
$-\delta$        & $+0.007$        & $+0.007$ \\
$-\theta$        & $-0.004$        & $+0.006$ \\
$-\alpha$        & $\mathbf{-0.021^{*}}$   & $\mathbf{-0.045^{***}}$ \\
$-\beta$         & $\mathbf{-0.008^{*}}$ & $\mathbf{-0.022^{***}}$ \\
$-\text{low }\gamma$    & $+0.003$        & $+0.001$ \\
$-\text{high }\gamma$   & $0.000$         & $0.000$ \\
\bottomrule
\end{tabular}
\end{table}

\subsection{Target-specific analyses reveal no detectable differences}

The target-specific analyses provided no evidence of differences among targets within either task. No target-dependent cluster survived correction in the repeated-measures spatiotemporal analysis of the target-specific task-versus-no-task maps. 
Likewise, none of the significant primary task-versus-no-task clusters in either task showed a target effect in the within-cluster Friedman analyses (speech: Kendall's $W = 0.010$--$0.079$, Holm-corrected $p = 0.482$--$1.000$; motor: $W = 0.022$--$0.037$, all Holm-corrected $p = 1.000$).

Pairwise FB-CSP decoding among the three imagery targets within each task provided no evidence of target-specific separability, with mean balanced accuracies of $0.476$--$0.487$ for speech imagery and $0.475$--$0.491$ for motor imagery. None exceeded chance after Holm correction
(one-sided Wilcoxon signed-rank tests against $0.5$). 

\subsection{Best-performing participants reveal focal alpha synchronization in right frontocentral region} 

\begin{figure}[t]
    \centering
    \subfloat[Early alpha activity for $6$ best-performing participants.
    \label{fig:best_performing_subset_alpha_power}]{%
        \makebox[\columnwidth][c]{%
            \includegraphics[width=0.7\columnwidth]{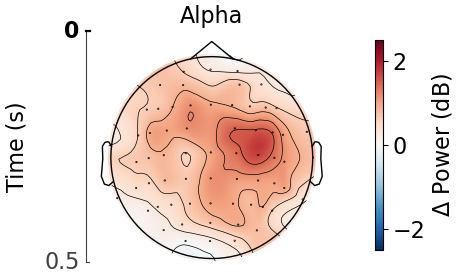}%
        }%
    }\\[1em]
    \subfloat[Time-resolved $10$~Hz power at FC4.\label{fig:time_resolved_alpha_power}]{%
        \includegraphics[width=\columnwidth]{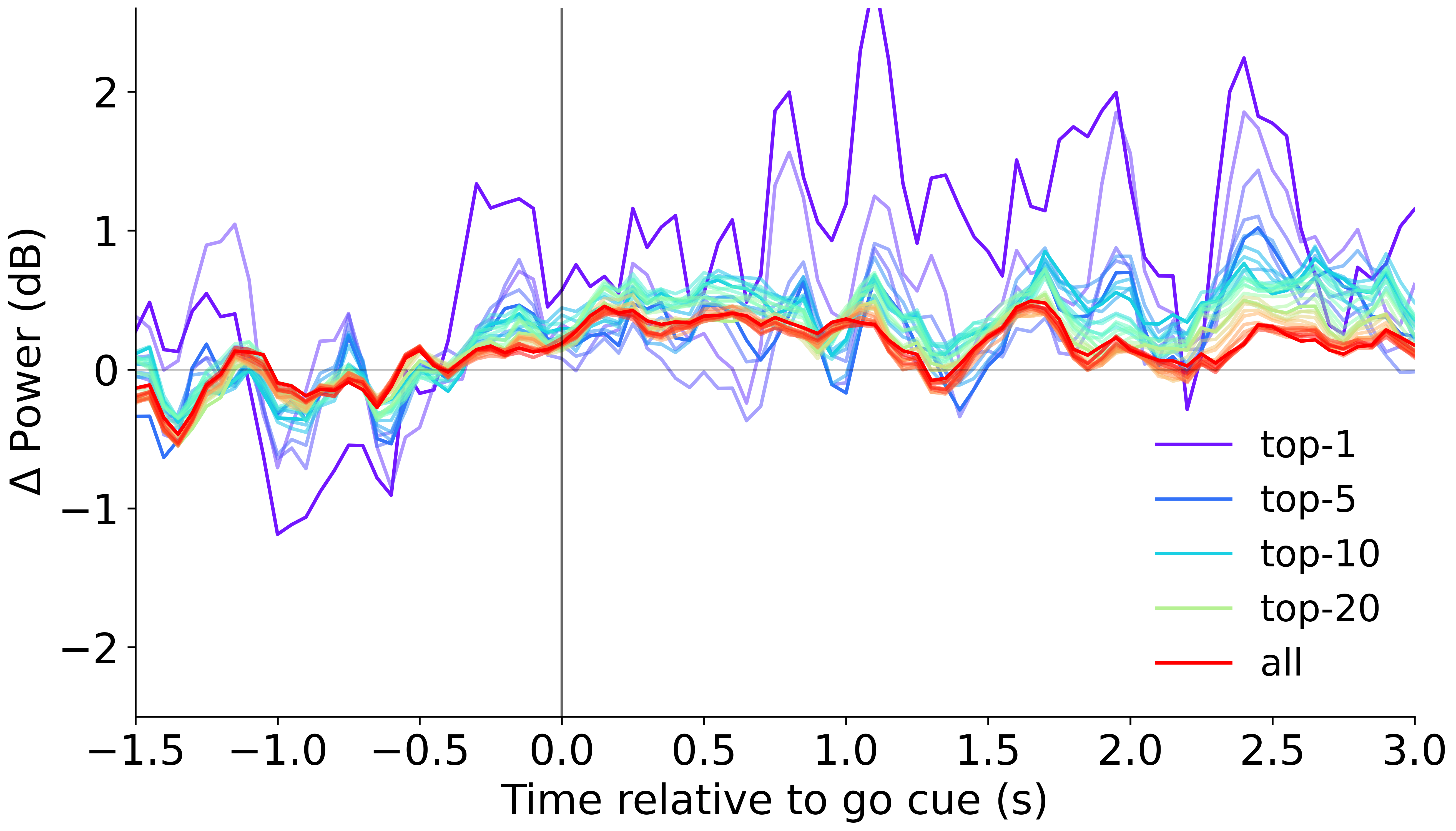}%
    }
    \caption{Alpha synchronization in the right frontocentral region for best-performing participants.}
    \label{fig:significant_speech_imagery}
\end{figure}

As a descriptive follow-up, we sought to reduce the influence of poorly performing participants. Using a binomial threshold for speech-imagery detection ($37/60$, balanced accuracy $0.617$), we retained $6$ participants (for reference, $28$ of $34$ participants had balanced accuracy $> 0.5$). Fig.~\ref{fig:best_performing_subset_alpha_power} shows the early alpha response for these best-performing participants and reveals that the right-frontocentral synchronization centered on FC4 was more pronounced in these participants.

We therefore used FC4 as the representative site for time-resolved analysis. Figure~\ref{fig:time_resolved_alpha_power} shows the $10$~Hz task-versus-no-task power difference extracted with a Morlet wavelet transform for the full cohort and for progressively smaller top-$k$ subsets ranked by FB-CSP speech-detection performance. 
Across the full cohort, the alpha difference was already present near cue onset and remained predominantly positive throughout the execution period and beyond. The time courses showed a shared transient dip around $1.2$--$1.4$~s, followed by a rebound. Higher-performing subsets generally showed larger positive amplitudes while retaining a broadly similar temporal profile.

\section{DISCUSSION}

Finger motor imagery showed the expected contralateral alpha and low-beta desynchronization over sensorimotor scalp, whereas speech imagery did not resemble this canonical pattern and instead showed weaker, more spatially distributed alpha synchronization.
Although the observed alpha synchronization is consistent with previous imagined-speech EEG findings \cite{Tsukahara2019}, our matched within-participant design characterizes it relative to both a no-task condition and a direct motor-imagery reference.

Although intracranial studies have reported low-beta decreases during imagined speech \cite{ProixEtAl2022}, the present scalp response was dominated by alpha synchronization, alongside weaker changes across the other frequency bands. Alpha increases have more generally been associated with inhibitory gating, suppression of task-irrelevant processing, and controlled access under demanding task conditions \cite{JensenMazaheri2010,FoxeSnyder2011,Klimesch2012}. Alpha activity can also lateralize over sensorimotor regions during top-down somatosensory attention and increase during motor inhibition \cite{Haegens2011,Bonstrupp2015}. These accounts offer possible interpretations of the dominant alpha response, although the present analyses cannot exclude weaker motor- or sensory-related effects that were obscured by broader task-related activity or varied across participants.


Baseline-normalized speech-imagery activity without no-task subtraction retained the dominant alpha synchronization but showed a somewhat different spatial distribution. Although the matched no-task contrast is arguably better suited to isolate task-related activity under the same trial structure than a temporally distant baseline, this comparison shows that the detailed spatial pattern is sensitive to the chosen reference and should be interpreted cautiously.

The association between speech-imagery and finger-motor-imagery detection performance suggests that both tasks may be influenced by common factors, including trial-to-trial consistency, strength of task-related neural signals, and EEG signal quality. Because the cohort was young and healthy with a narrow age range, the present data provide little basis for assessing demographic or clinical predictors.

The absence of detectable target effects in both the spatiotemporal patterns and pairwise decoding suggests that the pooled task-versus-no-task signatures were not evidently driven by a single speech command or finger movement. This does not establish equivalence among targets, however, and subtle, temporally variable, or spatially focal class-specific differences may remain below the sensitivity of the present analyses.

Our results do not uniquely determine the processes underlying the observed speech-imagery pattern. At the group level, however, this pattern is more consistent with broader task-related processes than with a straightforward articulatory counterpart of canonical motor imagery.

\section{CONCLUSION}

This study provides a group-level characterization of speech imagery in scalp EEG using a matched within-participant design comprising speech imagery, finger motor imagery, and no-task trials. Finger motor imagery showed the expected contralateral alpha and low-beta desynchronization over left sensorimotor scalp, whereas speech imagery showed weaker and more spatially distributed alpha synchronization. Even when participants were instructed to imagine articulatory movements, the dominant scalp pattern did not resemble canonical motor imagery, suggesting that speech imagery in scalp EEG is shaped more by broader task-related processes, potentially including attention, than by a straightforward articulatory-motor interpretation.





\section*{ACKNOWLEDGMENT}

This work was supported in part by Horizon Europe’s Marie
Sk{\l}odowska-Curie Action (No. 101118964), Horizon 2020 research
and innovation programme (No. 857375), the special research fund
of KU Leuven (C24/18/098), the Belgian Fund for Scientific Research
-- Flanders (1S65622N, G0A4118N, G0A4321N, G0C1522N), and the Hercules
Foundation (AKUL 043), and the China Scholarship Council. 
Corresponding author: Bob Van Dyck (e-mail: bob.vandyck@kuleuven.be).

\bibliographystyle{ieeetr}
\bibliography{references}

\end{document}